\UseRawInputEncoding
\documentclass[aps,prb,reprint,
]{revtex4-2}
\usepackage{amsthm}
\usepackage{mathrsfs}
\usepackage{amssymb}
\usepackage{epsfig}
\usepackage{amsmath}
\usepackage{graphicx}
\usepackage{epsfig}
\usepackage{booktabs}
\usepackage{multirow}
\usepackage[colorlinks,urlcolor=blue]{hyperref}
\usepackage{makecell}
\usepackage{bm}
\usepackage[utf8]{inputenc}
\usepackage{lipsum}
\usepackage{soul}
\soulregister\cite7

\usepackage{dcolumn}
\usepackage{amsthm}
\usepackage{color}
\newtheoremstyle{query}%
{}{}
{\color{red}}
{}
{\sffamily\bfseries}{:}{12pt}
{}
\theoremstyle{query}
\newtheorem{aq}{Author Query/Comment}

\newcommand{\baq}{\begin{aq}}
\newcommand{\eaq}{\end{aq}}

\begin{document}
\title{Strain-tunable Dzyaloshinskii--Moriya interaction and skyrmions in two-dimensional Janus Cr$_{2}$X$_{3}$Y$_{3}$ (X, Y = Cl, Br, I, X $\neq$ Y) trihalide monolayers}

\author{Zhong Shen}

\author{Changsheng Song}
\email[]{cssong@zstu.edu.cn}

\author{Yufei Xue}
\author{Zebin Wu}
\affiliation{Key Laboratory of Optical Field Manipulation of Zhejiang Province, Department of Physics, Zhejiang Sci-Tech University, Hangzhou 310018, China}
\author{Jiqing Wang}
\email[]{jqwang@ee.ecnu.edu.cn}
\address{School of Physics and Electronic Science, East China Normal University, Shanghai 200241, China}
\author{Zhicheng Zhong}
\email[]{zhong@nimte.ac.cn}
\address{Key Laboratory of Magnetic Materials and Devices, Zhejiang Province Key Laboratory of Magnetic Materials and Application Technology, Ningbo Institute of Materials Technology and Engineering, Chinese Academy of Sciences, Ningbo 315201, China}
\address{China Center of Materials Science and Optoelectronics Engineering, University of Chinese Academy of Sciences, Beijing 100049, China}



\date{\today}

\begin{abstract}	
Recently, great effort has been devoted to the search for two-dimensional (2D) ferromagnetic materials with inherent strong Dzyaloshinskii--Moriya interaction (DMI).
Here, through a first-principles approach, we systematically investigate the effect of biaxial strain on the DMI, the Heisenberg exchange interaction, and the magnetic anisotropy energy (MAE) of Janus Cr$_{2}$X$_{3}$Y$_{3}$ (X, Y = Cl, Br, I, X $\neq$ Y)  monolayers. Both DMI and MAE can be significantly enhanced by tensile strain, while a reversal of the chirality of DMI in Cr$_{2}$Cl$_{3}$Br$_{3}$ and a switch of MAE from off-plane to in-plane in Cr$_{2}$I$_{3}$Cl$_{3}$ are induced by a compressive strain of $2\%$. Microscopically, DMI and MAE are  associated mainly with the large spin--orbit coupling of the heavy nonmagnetic halogen atoms rather than that of the magnetic Cr atoms. In particular, the peculiar magnetic transition of Cr$_{2}$I$_{3}$Cl$_{3}$ is explained by  competition between direct exchange and superexchange interactions. Micromagnetic simulations show that  a small external magnetic field of 65~mT stabilizes a skyrmion with a diameter of 9.8~nm in the Cr$_2$I$_3$Cl$_3$ monolayer. Our results will provide guidance for further research on DMI and skyrmions in 2D Janus materials, as well as a basis for the potential applications in spintronic devices.
\end{abstract}
\maketitle
\section{Introduction}
\label{sec:I}

Skyrmions are quasiparticles with unique topological protection characteristics and a vortex-like noncollinear spin texture with  opposite directions between the central and edge spins~\cite{RN2009,RN120,RN121,RN109}.
Driven by a very small spin-polarized current, skyrmions can move accompanied by the skyrmion Hall effect~\cite{RN120,RN1987,RN1714}.
These peculiar phenomena enable skyrmions to be used in a variety of new types of spintronic devices~\cite{RN116}, such as track memory~\cite{RN887,RN1986,RN1988}, microwave detectors~\cite{RN1989,RN1990}, and nano-oscillators~\cite{RN1991,RN1993}, among which  track memory is expected to provide the next generation of nonvolatile, high-performance, low-energy-consumption, and high-density memory devices. The emergence and stability of magnetic skyrmions are mostly determined by the Dzyaloshinskii--Moriya interaction (DMI)~\cite{RN121}, which is essentially a magnetic interaction caused by spin--orbit coupling (SOC) and breaking of spatial inversion symmetry.

Magnetic skyrmions were first found in single-crystal materials with $B20$ structure, such as MnGe~\cite{RN970} and MnSi~\cite{RN1994}. Later, it was discovered that magnetic skyrmions can exist in the more easily manipulated  ultrathin metal layers with broken spatial inversion symmetry, such as Fe/Ir(111)~\cite{RN118} and PdFe/Ir(111)~\cite{RN3710}.
Since then, great effort has been made to calculate the magnetic parameters~\cite{RN3695,RN3099,RN33} and to grow ferromagnet/heavy metal (FM/HM) heterostructures, such as Pt/Co/MgO~\cite{RN1996} and Ir/Fe/Co/Pt~\cite{RN1995}, whose interface DMI and off-plane magnetic anisotropy can be artificially adjusted by changing the film thickness, material combination, and other parameters~\cite{RN121}.

However, for FM/HM heterostructures, their interfacial defects and atomic stacking order strongly affects the facial DMI~\cite{RN1995} but are hard to control experimentally. Therefore, searching for and investigating 2D magnetic materials with intrinsic strong DMI is of great importance. Nonetheless, many 2D magnets are centrosymmetric with an absence of DMI, examples being CrI$_{3}$~\cite{RN11,RN3102} and Cr$_{2}$Ge$_{2}$Te$_{6}$~\cite{RN1037}.
Recently, a class of Janus materials with inherent broken spatial inversion symmetry have been reported to have strong SOC~\cite{PhysRevB.101.184401,RN31} and large DMI that can stabilize skyrmions~\cite{RN1042,RN1059}. 

Inspired by this, on the basis of the experimentally synthesized CrI$_{3}$ monolayer, we replace I atoms with Cl or Br to form Janus structures with intrinsic space inversion asymmetry.  We demonstrate by first-principles calculations that the DMI and  magnetic anisotropy
energy (MAE) of Cr$_{2}$X$_{3}$Y$_{3}$ (X, Y = Cl, Br, I, X $\neq$ Y) monolayers can be greatly enhanced by the application of tensile strain. In particular, with a compressive strain of $-2\%$, a chirality reversal of DMI and a switch of MAE from off-plane to in-plane appear in Cr$_{2}$Cl$_{3}$Br$_{3}$ and Cr$_{2}$I$_{3}$Cl$_{3}$, respectively. Microscopically, the strain-tunable DMI and MAE can be attributed to the strong SOC induced by the heavy nonmagnetic halogen atoms. Besides, the peculiar magnetic transition of Cr$_{2}$I$_{3}$Cl$_{3}$ is explained by the competition between direct exchange and superexchange interactions. In addition, micromagnetic simulations are performed with the magnetic parameters obtained from first-principles calculations, and a stable skyrmion with sub-10~nm diameter is found in an unstrained Cr$_{2}$I$_{3}$Cl$_{3}$ monolayer. Our work provides references for the study of skyrmions and DMI in 2D Janus monolayers, as well as guidance for spintronic applications.

\section{Calculational Methods}
\label{II}

\subsection{First-principles calculations and micromagnetic simulation}
We use the framework of density-functional theory (DFT) as implemented in the Vienna Ab initio Simulation Package (VASP)~\cite{RN2004} to perform our first-principles calculations on Cr$_{2}$X$_{3}$Y$_{3}$ (X, Y = Cl, Br, I, X $\neq$ Y) monolayers, with the projected augmented wave (PAW) method~\cite{RN1060,RN1064,RN1061}  describing the electron--core interaction. The generalized gradient approximation (GGA) of Perdew--Burke--Ernzerhof (PBE)~\cite{RN1999} is chosen to treat the exchange correlation effects, with an effective Hubbard-like term $U = 3$~eV for $3d$ electrons of Cr~\cite{RN113,RN127,RN34}. The plane-wave cutoff energy is set as 500~eV,
and the first Brillouin-zone integration is carried out using  $9 \times 9 \times 1$ and $5 \times 10 \times 1$ $\Gamma$-centered $k$-point meshes for the primitive cell and the $2\times1\times1$ supercell, respectively. To obtain accurate DMI parameters, we set a high convergence standard, with the energy and force less than $10^{-6}$~eV and 0.001~eV/\AA, respectively.
 
To explore the spin textures of the Cr$_{2}$X$_{3}$Y$_{3}$ system, we perform micromagnetic simulation with the Heisenberg model and Landau--Lifshitz--Gilbert (LLG) equation~\cite{RN1984,RN1985} as implemented in the \textit{Spirit} package~\cite{RN1055}. A $70\times70\times1$ supercell with periodic boundary conditions is chosen, and the number of iterations is set to $2\times10^{5}$ to reach the stable state at each temperature point.

\subsection{Dzyaloshinskii--Moriya interaction}
To obtain the DMI strength, the chirality-dependent total energy difference approach~\cite{RN33,RN32,RN31,RN30} is used, which has been successfully employed for DMI calculations in frustrated bulk systems and insulating chiral-lattice magnets, as well as in 2D Janus materials~\cite{RN30}. 
We take a
Hamiltonian with the following form~\cite{RN17,RN30,RN31}:
\begin{align}\label{1}
H = {}&\sum_{ \langle i,j  \rangle}J_{1}(\vec{S_{i}}\cdot\vec{S_{j}}) + \sum_{ \langle i,k  \rangle}J_{2}(\vec{S_{i}}\cdot\vec{S_{k}}) + \sum_{ \langle i,l  \rangle}J_{3}(\vec{S_{i}}\cdot\vec{S_{l}}) \nonumber \\
&+ \sum_{ \langle i,j  \rangle}\vec{d}_{ij}\cdot(\vec{S_{i}}\times\vec{S_{j}}) + 
\sum_{i}K(S_{i}^{z})^2.
\end{align}
Here, $J_{1}$, $J_{2}$, and $J_{3}$ are the Heisenberg exchange coefficients between nearest-neighbor, second-nearest-neighbor, and third-nearest-neighbor Cr atoms, and $\vec{d}_{ij}$ is the DMI vector between spins $\vec S_{i}$ and $\vec S_{j}$. $K$ is the single-ion anisotropy coefficient,  $\vec{S_{i}}$ and $\vec{S_{j}}$  are the spins of the $i$ and $j$ sites, and $S_{i}^{z}$ is the $z$ component of the spin at the $i$ site.

As summarized by Moriya~\cite{RN131}, if there is a mirror plane perpendicular to the Cr--Cr bond and passing through the middle of the bond, then the  DMI vector $\vec{d}_{ij}$ between the nearest-neighbor Cr atoms is in the mirror plane with the form 
\begin{equation}\label{2}
\vec{d}_{ij} = d_{\|}(\vec{u}_{ij}\times\vec{z}) + d_{z}\vec{z},
\end{equation}
where $\vec{u}_{ij}$ and $\vec{z}$ are the unit vectors from site $i$ to site $j$ and pointing along the $z$ direction, respectively. Approximately, $\vec{d}_{ij} = \vec{d}_{1} -  \vec{d}_{2}$~\cite{RN30,Frederic1962}, where $\vec{d}_{1}$, $\vec{d}_{2}$ will be explained in detail later [see Fig.~\ref{fig1}(b)].

To evaluate the in-plane component $d_{\|}$, we choose the clockwise (CW) and anticlockwise (ACW)  spin configurations with opposite chirality, as shown in Fig.~\ref{fig1}(a) by yellow arrows. 
$d_{\|}$ can be obtained from the following formula (a detailed derivation can be found in the Supplemental Material~\cite{suppl_mater}): 
\begin{equation}\label{3}
d_{\|} = -\frac{E_\mathrm{CW} - E_\mathrm{ACW}}{4\sqrt{3}S^2},
\end{equation}
where $E_\mathrm{CW}$ and $E_\mathrm{ACW}$ are the energies of Cr$_{2}$X$_{3}$Y$_{3}$ monolayers with CW and ACW spin configurations, respectively, and $S$ is the normalized spin~\cite{RN33}.
To explore the microscopic physical mechanism of DMI, we calculate the SOC energy difference~\cite{PhysRevB.88.184423,RN33} $\Delta E_\mathrm{SOC}$ of Cr$_2$X$_3$Y$_3$ to see the contributions to DMI from different atoms.  $\Delta E_\mathrm{SOC}$ is extracted from the  self-consistent total energy calculations of different spin configurations with opposite chirality when SOC is included.

\section{Results and discussion}

\begin{figure*}[!t] 
		\includegraphics[width=14.0cm]{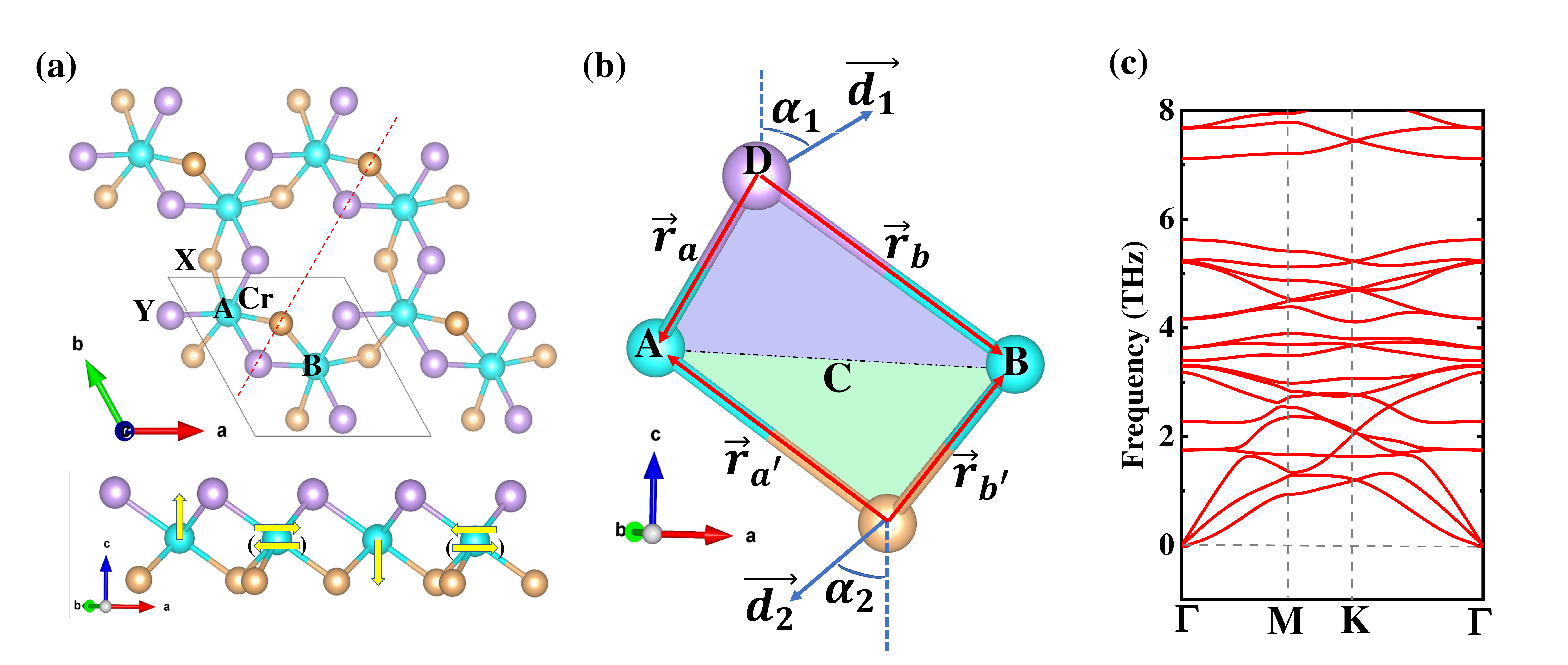}
		\caption{\label{fig1}
		(a) Top and side views of  Cr$_2$X$_3$Y$_3$ (X, Y = Cl, Br, I, X $\neq$ Y) monolayers. The solid lines in the top view show the primitive cell. $A$ and $B$ are the selected reference atoms, with the red dotted line being the mirror plane perpendicular to $AB$. The yellow vectors in the bottom view demonstrate the spin configurations to obtain the in-plane DMI component. (b) Schematic  of $\vec{d}_1$ and $\vec{d}_2$. $D$ is a nonmagnetic halogen atom with strong SOC. $\vec{r}_a$, $\vec{r}_{a'}$, $\vec{r}_b$, and $\vec{r}_{b'}$ are vectors from halogen atoms pointing to $A$ or $B$, as shown by the red arrows. $\alpha_1$ and $\alpha_2$ are the angles between $\vec{d}_1$ and the $z$ direction and between $\vec{d}_2$ and the $z$ direction. (c) Phonon dispersion spectrum of Cr$_2$I$_3$Cl$_3$ along the high-symmetry points in reciprocal space.
			 }
\end{figure*}

Top and side views of Janus Cr$_2$X$_3$Y$_3$ monolayers are shown in Fig.~\ref{fig1}(a). The Cr atoms form a honeycomb network sandwiched by two nonmagnetic atomic planes with different halogen atoms X and Y.
We now give a detailed description of the direction of the DMI. There are five rules to determine the direction of the DMI between two magnetic atoms $A$ and $B$, with $C$ being the midpoint of $A$ and $B$, which are summarized  by Moriya~\cite{RN131}. 
From Figs.~\ref{fig1}(a) and~\ref{fig1}(b), we can see that the symmetry of {Cr}$_2$X$_3$Y$_3$ satisfies the second  of Moriya's  symmetry rules: 
when a mirror plane perpendicular to $AB$ passes through $C$, $\vec{d}\parallel$\,mirror plane or $\vec{d}\,\bot\,AB$.
However, this does not give us a precise direction of the DMI. Thus, from the further discussion in~\cite{Frederic1962}, we can determine the direction of $\vec{d}_1$ along $\vec{r}_b \times \vec{r}_a$ and $\vec{d}_2$ along $\vec{r}_{b'} \times \vec{r}_{a'}$, with $\vec{d}=\vec{d}_1-\vec{d}_2$.
Phonon dispersions are calculated to examine the stability of Cr$_2$X$_3$Y$_3$ in Fig.~\ref{fig1}(c) and Fig.~S1 (Supplemental  Material~\cite{suppl_mater}).  There is no imaginary frequency in the case of Cr$_2$I$_3$Cl$_3$, which suggests that  it is dynamically stable. The small imaginary frequencies of Cr$_2$Cl$_3$Br$_3$ and Cr$_2$I$_3$Br$_3$ indicate  slight structural instability, which is often encountered in the phonon spectra of other 2D materials~\cite{RN30,RN1073}.

\begin{table*}[!t]
	\caption{\label{table1}
		Optimized lattice constants $a$, angular difference $\Delta \theta$ between Cr--X--Cr ($\theta_1$) and Cr--Y--Cr ($\theta_2$), bond length difference $\Delta r$ between Cr--X ($r_1$) and Cr--Y ($r_2$) [as shown in Fig.~\ref{fig2}(e)], Heisenberg exchange coefficients $J_1$, $J_2$, $J_3$, and magnetic moment $M_\mathrm{Cr}$ of Cr atoms in Cr$_2$X$_3$Y$_3$ monolayers with different strains.}
		\begin{ruledtabular}
	\begin{tabular}{ldddddddd}
			& \multicolumn{1}{c}{Strain (\%)} & \multicolumn{1}{c}{$a$ (\AA)}  & \multicolumn{1}{c}{$\Delta \theta$ (deg)} & \multicolumn{1}{c}{$\Delta r$ (\AA)} & \multicolumn{1}{c}{$J_{1}$ (meV)} & \multicolumn{1}{c}{$J_2$ (meV)} & \multicolumn{1}{c}{$J_3$ (meV)} & \multicolumn{1}{c}{$M_\mathrm{Cr}$ ($\mu_B$)} \\
			\hline
			\multirow{6}[2]{*}{Cr$_2$Cl$_3$Br$_3$} & -2    & 6.11  & 9.21  & 0.17  & -4.77  & -0.62  & 0.16  & 3.01  \\
				  & 0     & 6.23  & 9.98  & 0.18  & -4.37  & -0.54  & 0.11  & 3.02  \\
				  & 2     & 6.36  & 10.62  & 0.19  & -3.21  & -0.47  & 0.07  & 3.04  \\
				  & 4     & 6.48  & 11.06  & 0.19  & -1.91  & -0.42  & 0.05  & 3.06  \\
				  & 6     & 6.61  & 11.42  & 0.19  & -0.86  & -0.38  & 0.03  & 3.08  \\
				  & 8     & 6.73  & 11.66  & 0.19  & -0.26  & -0.35  & 0.02  & 3.11  \\
				  \hline
			\multirow{6}[2]{*}{Cr$_2$I$_3$Br$_3$} & -2    & 6.60  & 10.23  & 0.20  & -5.50  & -1.25  & 0.02  & 3.14  \\
				  & 0     & 6.73  & 10.77  & 0.21  & -4.79  & -1.11  & -0.01  & 3.16  \\
				  & 2     & 6.87  & 11.04  & 0.21  & -3.98  & -1.00  & -0.02  & 3.19  \\
				  & 4     & 7.00  & 11.16  & 0.21  & -3.46  & -0.91  & -0.02  & 3.22  \\
				  & 6     & 7.14  & 11.19  & 0.20  & -3.41  & -0.84  & -0.02  & 3.26  \\
				  & 8     & 7.27  & 10.10  & 0.18  & -3.92  & -0.78  & -0.02  & 3.30  \\
				  \hline
			\multirow{6}[2]{*}{Cr$_2$I$_3$Cl$_3$} & -2    & 6.42  & 20.57  & 0.39  & -2.62  & -1.26  & -0.20  & 3.10  \\
				  & 0     & 6.55  & 21.47  & 0.39  & -1.48  & -1.09  & -0.19  & 3.12  \\
				  & 2     & 6.68  & 22.01  & 0.39  & -0.45  & -0.96  & -0.17  & 3.15  \\
				  & 4     & 6.81  & 22.22  & 0.39  & 0.17  & -0.87  & -0.14  & 3.18  \\
				  & 6     & 6.94  & 22.01  & 0.38  & 0.16  & -0.80  & -0.12  & 3.22  \\
				  & 8     & 7.08  & 21.31  & 0.36  & -0.49  & -0.75  & -0.10  & 3.26  
	  \end{tabular}%
	  \end{ruledtabular}
	\label{tab:addlabel}%
  \end{table*}%

As shown in Table~\ref{table1},  the relaxed lattice constants $a$ of Cr$_2$Cl$_3$Br$_3$, Cr$_2$I$_3$Cl$_3$, and Cr$_2$I$_3$Br$_3$ are 6.23, 6.55, and 6.73~\AA,  respectively, and they increase as functions of the radius of the nonmagnetic atoms X (Y). 
In our work, we do not further investigate the magnetic behaviors of Cr$_{2}$X$_{3}$Y$_{3}$ with  compressive strain greater than 2$\%$, because the DMI tends to become weaker as the compressive strain increases, which is not conducive to the creation of skyrmions.

\begin{figure*}[!t]
		\includegraphics[width=14.0cm]{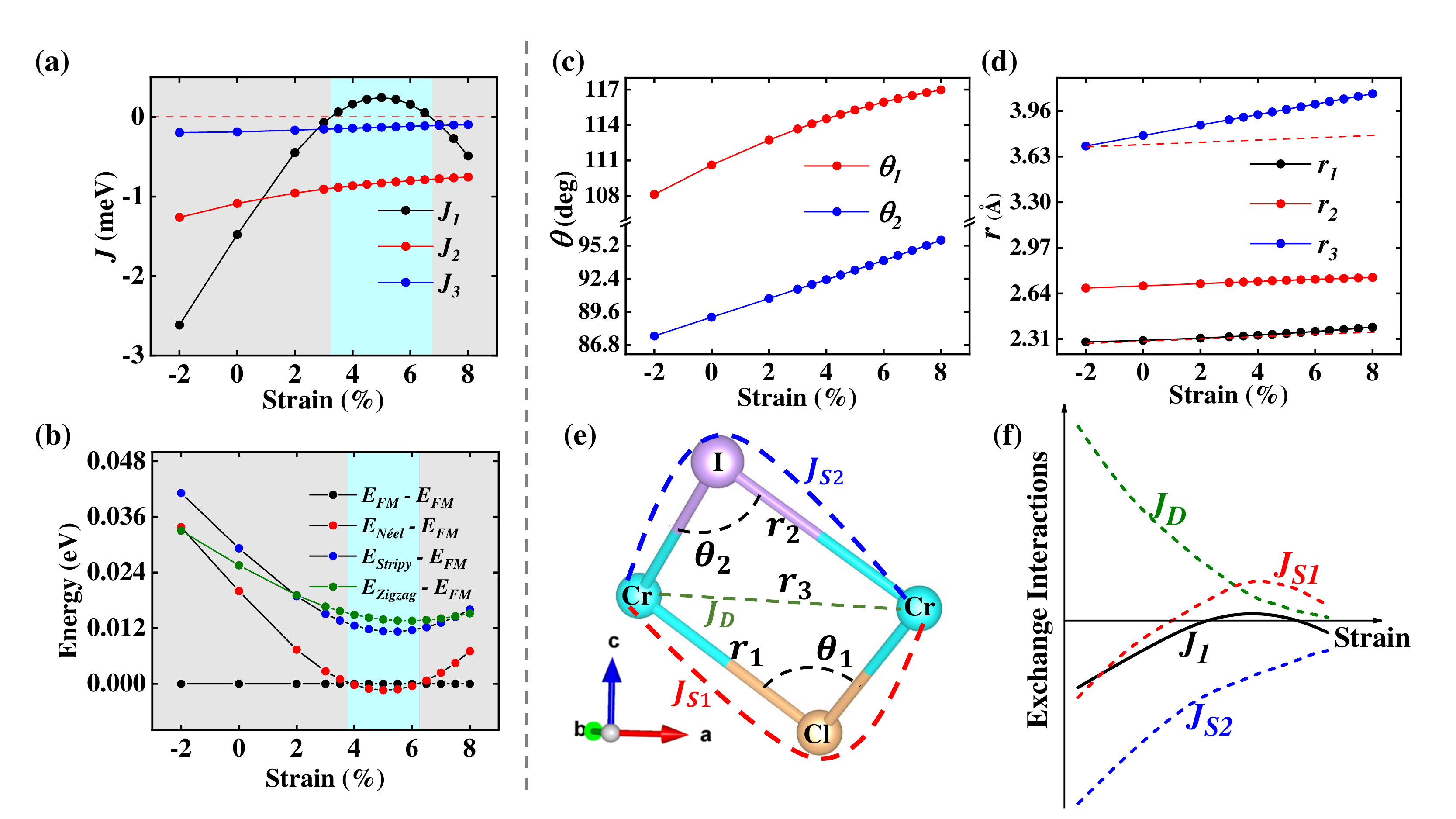}
		\caption{\label{fig2}
(a) Heisenberg exchange coefficients $J_1$, $J_2$,  $J_3$, (b) energies of the four spin configurations (\textit{FM}, \textit{N\'eel}, \textit{Stripy}, and \textit{Zigzag}), (c) bond angles of Cr--Cl--Cr ($\theta_1$) and Cr--I--Cr ($\theta_2$), and (d) bond lengths of Cr--Cl ($r_1$), Cr--I ($r_2$), and Cr--Cr ($r_3$) as functions of biaxial strain. The red dashed line in (d) illustrates the slope of $r_2$. (e) Schematic  of $\theta_1$, $\theta_2$, $r_1$, $r_2$, $r_3$, superexchange $J_{S1}$ and $J_{S2}$, and direct exchange $J_D$. (f) Schematic  of Heisenberg exchange interactions as functions of biaxial strain.}
\end{figure*}

We also notice that $J_1$ and $J_2$ of both Cr$_2$Cl$_3$Br$_3$ and Cr$_2$I$_3$Br$_3$ are negative in the range of $-2\%$ to $8\%$ biaxial strain, as shown in Table~\ref{table1}, while $J_3$ plays a less important role because it is more than an order of magnitude smaller than $J_1$. According to Eq.~\eqref{1}, a negative value of the Heisenberg exchange coefficient corresponds to ferromagnetic (FM) coupling, while a positive value corresponds to antiferromagnetic (AFM) coupling. Thus, the ground states of Cr$_2$Cl$_3$Br$_3$ and Cr$_2$I$_3$Br$_3$ are FM in the range of biaxial strain from $-2\%$ to $8\%$. For the Cr$_2$I$_3$Cl$_3$ monolayer, we find that $J_1$ changes  sign twice when tensile strain is applied. To understand this peculiar behavior, we extract the energies of four spin configurations (\textit{FM}, \textit{N\'eel}, \textit{Stripy}, and \textit{Zigzag}) and calculate the bond angles of Cr--Cl--Cr ($\theta_1$), Cr--I--Cr ($\theta_2$) and the bond lengths of Cr--Cl ($r_1$), Cr--I ($r_2$), and Cr--Cr ($r_3$), as shown in Fig.~\ref{fig2}. The system favors the FM configuration in the range of $-2\%$ to 3.5$\%$ biaxial strain. However, a transition from FM to AFM occurs at a strain of 4$\%$, with the energy of the \textit{N\'eel} AFM configuration being the lowest, as shown by the cyan area (II) in Fig.~\ref{fig2}(b). Then, with   increasing strain, another transition from AFM to FM appears at a strain of 6.5$\%$, after which the system favors the FM configuration again. We notice that the trend of the change in energy  of the \textit{N\'eel} AFM configuration is similar to that of $J_1$. We further explore the microscopic physical mechanisms of the transitions of $J_1$ using  direct exchange and superexchange interaction theory~\cite{Goodenough1955,Kanamori1959,Anderson1959}. As shown in Fig.~\ref{fig2}(e),
$J_1$ can be expressed as $J_1 = J_{S1}+J_{S2}+J_D$, where $J_{S1}$ and $J_{S2}$ represent the superexchange interactions of the paths Cr--I--Cr and Cr--Cl--Cr, and $J_D$ represents the direct exchange interaction of Cr--Cr. Then, it can be seen from Fig.~\ref{fig2}(c) that $\theta_1$ and $\theta_2$ are both close to $90^\circ$ when the tensile strain is small. Thus, according to the Goodenough--Kanamori--Anderson (GKA) rules~\cite{Goodenough1955,Kanamori1959,Anderson1959},  both of the superexchange interactions $J_{S1}$ and $J_{S2}$ are FM ($J_{S1}, J_{S2} < 0$), whereas the direct exchange interaction $J_D$ between nearest-neighbor Cr atoms is AFM ($J_D > 0$), and thus $J_1 = J_{S1}+J_{S2}+J_D$ has a negative value, corresponding to an FM coupling. Then, as the strain increases, $\theta_1$ deviates from $90^\circ$, causing the FM superexchange interaction $J_{S1}$ to become weak. A transition of $J_{S1}$ from FM to AFM occurs when the strain  increases further, as shown by the red dashed line in Fig.~\ref{fig2}(f). This transition is caused mainly by the AFM $J_{S1}$ and $J_D$, which compete with the FM $J_{S2}$. Then, a second transition of $J_1$ from AFM to FM occurs,   mainly as a result of the AFM $J_{S1}$ and $J_D$ decaying more rapidly than   the FM $J_{S2}$. This can be attributed to the more rapid increases  in $r_1$ and $r_3$ compared with $r_2$, as shown in Fig.~\ref{fig2}(d).

\begin{figure*}[!t]
		\includegraphics[width=16.0cm]{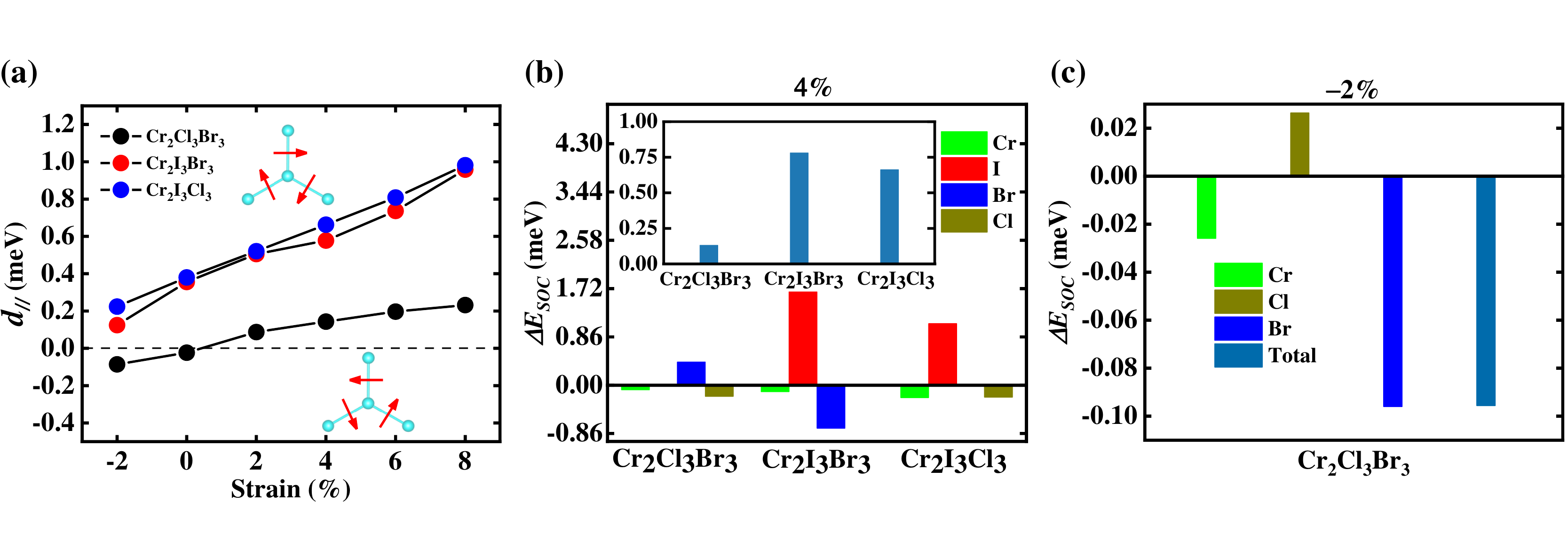}
		\caption{\label{fig3}
		(a) In-plane DMI component $d_{\|}$ of Cr$_2$X$_3$Y$_3$ monolayers as a function of biaxial strain. The inset shows the $d_{\|}$ (red arrow) between  nearest-neighbor Cr atoms. (b) Atom-resolved localization of the SOC energy difference $\Delta$$E_\mathrm{SOC}$ for Cr$_2$X$_3$Y$_3$ monolayers calculated with opposite chiralities under $4\%$ biaxial strain and (c) that of Cr$_2$Cl$_3$Br$_3$ under $-2\%$ biaxial strain. The total SOC energy variation in each material is shown in the inset in (b).
		}
\end{figure*}

From the differences in bond angles ($\Delta\theta$) and bond lengths ($\Delta r$) in Table~\ref{table1}, we can clearly see that the spatial inversion symmetry of Janus Cr$_2$X$_3$Y$_3$ is broken in the off-plane direction. A large DMI can then be induced by the broken spatial inversion symmetry, as shown in Fig.~\ref{fig3}(a).
Here, we focus mainly on the in-plane component $d_{\|}$ of the DMI, which plays a leading role compared with the off-plane component $d_{z}$ (a comparison between $d_{\|}$ and $d_{z}$ can be found in Table~S2 in the Supplemental Material~\cite{suppl_mater}).
As illustrated in Fig.~\ref{fig3}(a), intrinsic large $d_{\|}$ values of 0.38 and 0.36~meV are found in unstrained Cr$_2$I$_3$Cl$_3$ and Cr$_2$I$_3$Br$_3$ monolayers, respectively. Then, with increasing biaxial strain, $d_{\|}$ can be significantly enhanced up to 0.99 and 0.96~meV, which are more than two times larger than the values in the absence of  strain. As shown by the inset in Fig.~\ref{fig3}(a), positive and negative values of $d_{\|}$ correspond respectively to  CW and ACW arrangements of the in-plane DMI components.

To further understand the origin of strong DMI in Cr$_2$X$_3$Y$_3$ monolayers, as shown in Fig.~\ref{fig3}(b), we calculate the DMI-associated SOC energy difference $\Delta  E_\mathrm{SOC}$~\cite{RN33,RN30} of different atoms in Cr$_2$X$_3$Y$_3$ at a tensile strain of 4$\%$. The inset in Fig.~\ref{fig3}(b) shows the algebraic sum of $\Delta E_\mathrm{SOC}$ for each Cr$_{2}$X$_{3}$Y$_{3}$ monolayer, which corresponds to the strength of $d_{\|}$. Strong DMI is mainly associated with the large $\Delta E_\mathrm{SOC}$ located on the heavy nonmagnetic halogen atom X or Y, which is similar to what occurs in Co/Pt~\cite{RN33} and MnXY~\cite{RN30} systems. Besides,
in Cr$_{2}$X$_{3}$Y$_{3}$ monolayers, the $\Delta E_\mathrm{SOC}$ of X and Y atoms have opposite signs, which can be explained by the Fert--Levy model~\cite{RN1043,RN33,RN30}. Owing to the strong SOC of the halogen atoms X and Y, when a polarized electron transfers between Cr atoms through the intermediate atom X or Y, the spin direction of the electron is tilted by spin--orbit scattering, which leads to a tilt of local spins on adjacent Cr atoms. This tilt of local spins has two possibilities, namely, CW  and ACW. When the local spins are tilted CW (ACW), $\Delta E_\mathrm{SOC} = E_\mathrm{SOC}^\mathrm{ACW} - E_\mathrm{SOC}^\mathrm{CW}>0$ ($\Delta E_\mathrm{SOC} = E_\mathrm{SOC}^\mathrm{ACW} - E_\mathrm{SOC}^\mathrm{CW}<0$). For halogen atoms on opposite sides of the Cr layer, the CW and ACW are also opposite, with the result that the $\Delta E_\mathrm{SOC}$ of the X and Y atoms have opposite signs. 
Figure~$\ref{fig3}$(c) shows the atom-resolved $\Delta E_\mathrm{SOC}$ values of Cr$_2$Cl$_3$Br$_3$ for a strain of $-2\%$. The results show that the $\Delta E_\mathrm{SOC}$ of the Br atom plays a leading role, resulting in a negative total $\Delta$$E_\mathrm{SOC}$, which induces a negative $d_{\|}$ in Cr$_2$Cl$_3$Br$_3$.

\begin{figure*}[!t]
\includegraphics[width=14.0cm]{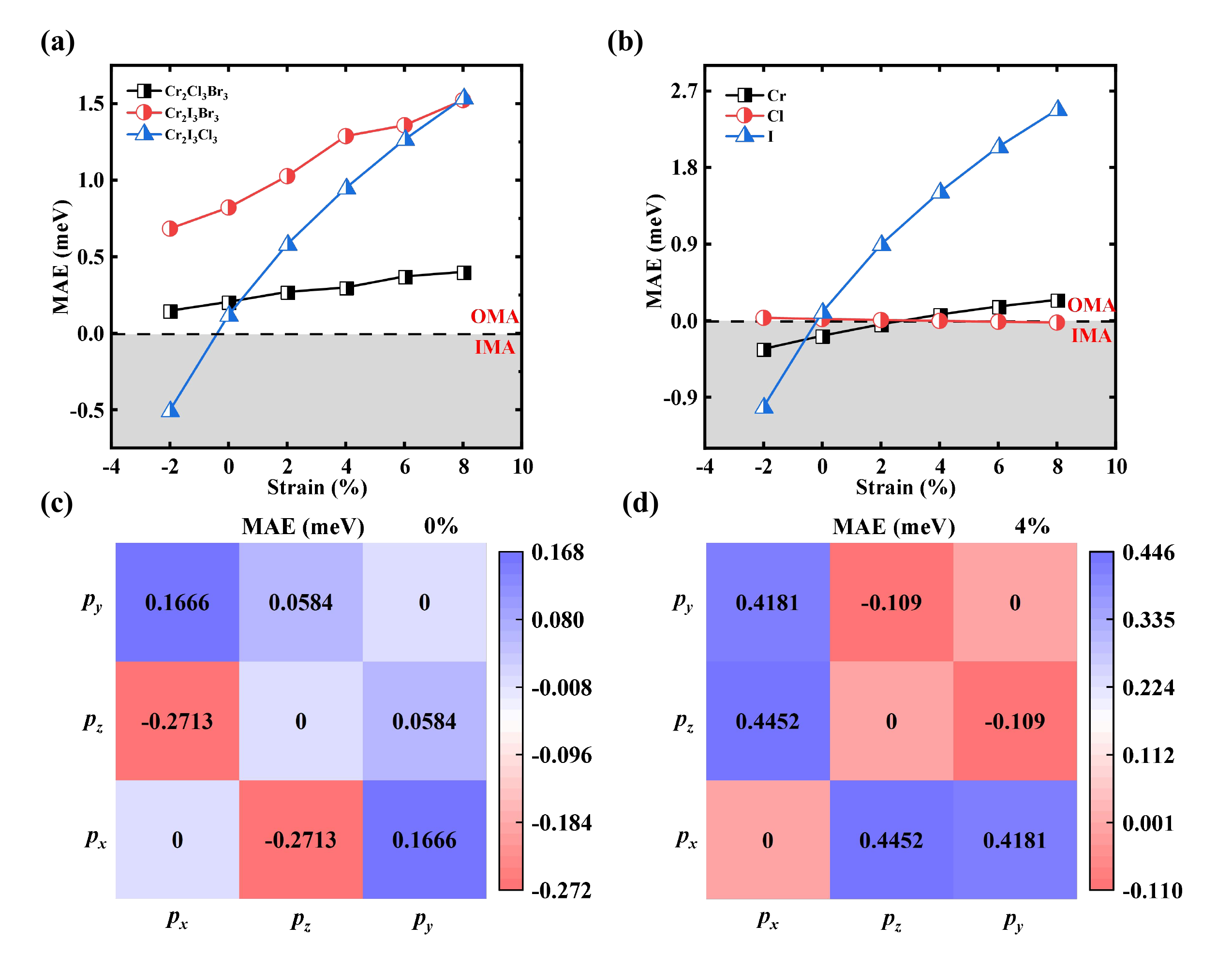}
\caption{\label{fig4}
 (a) Total MAE of Janus Cr$_2$X$_3$Y$_3$ monolayers and (b) atom-resolved MAE of Cr$_2$I$_3$Cl$_3$ as functions of biaxial strain. (c) and (d) Orbit-resolved MAE of the $5p$ orbitals of the I atom in Cr$_{2}$I$_{3}$Cl$_{3}$ with $0\%$ and $4\%$ biaxial strain, respectively. 
}
\end{figure*}

Not only do the DMI and Heisenberg exchange interaction  affect the magnetic structure of 2D ferromagnets, but also the magnetic anisotropy plays a vital role in the formation and stability of skyrmions. As we know, both the DMI and magnetic anisotropy  originate from SOC. The magnetic anisotropy energy (MAE) is defined as the energy difference between in-plane ($E_{x}$) and off-plane ($E_{z}$) FM states: MAE $= E_{x} - E_{z}$, with MAE $> 0$ and MAE $<0$ corresponding to off-plane and in-plane magnetic anisotropy (OMA and IMA), respectively.
As shown in Fig.~\ref{fig4}(a), the MAEs of Cr$_{2}$I$_{3}$Br$_{3}$ and Cr$_{2}$Cl$_{3}$Br$_{3}$ are positive and gradually increase as the strain changes from $-2\%$ to $8\%$. Interestingly, for the Cr$_2$I$_3$Cl$_3$ monolayer, there is a switch from OMA to IMA at a compression strain of $-2\%$.
To explore the microscopic physical mechanisms of MAE, we calculate the atom-resolved MAE of Cr$_{2}$I$_{3}$Cl$_{3}$ as a function of biaxial strain. As shown in Fig.~\ref{fig4}(b), the heavy nonmagnetic atom I rather than the magnetic atom Cr makes the most significant contribution to MAE. A similar phenomenon has been found in CrI$_{3}$~\cite{RN12,RN3540} and CrXTe (X = S, Se)~\cite{RN31} systems.
Furthermore, we calculate the orbit-resolved MAE of the $p$ orbitals of the I atom in Cr$_{2}$I$_{3}$Cl$_{3}$ at strains of $0\%$ [Fig.~\ref{fig4}(c)] and $4\%$ [Fig.~\ref{fig4}(d)]. At a strain of $0\%$, the hybridization between $p_{x}$ and $p_{y}$ orbitals contributes to positive MAE (OMA), while the hybridized $p_{x}$ and $p_{z}$ orbitals contribute to negative MAE (IMA). The competition between the hybridized $p_{x}$--$p_{y}$ and $p_{x}$--$p_{z}$ leads to the small OMA (MAE $> 0$) of Cr$_{2}$I$_{3}$Cl$_{3}$ at a strain of $0\%$. In the case of a strain of $4\%$, both the hybridization of $p_{x}$--$p_{y}$ and that of $p_{x}$--$p_{z}$ contribute to OMA, and the OMA contribution from $p_{x}$ and $p_{y}$ hybridization is enhanced more than twofold, which is responsible for the large OMA at a strain of $4\%$.

\begin{figure*}[!t]
		\includegraphics[width=14.0cm]{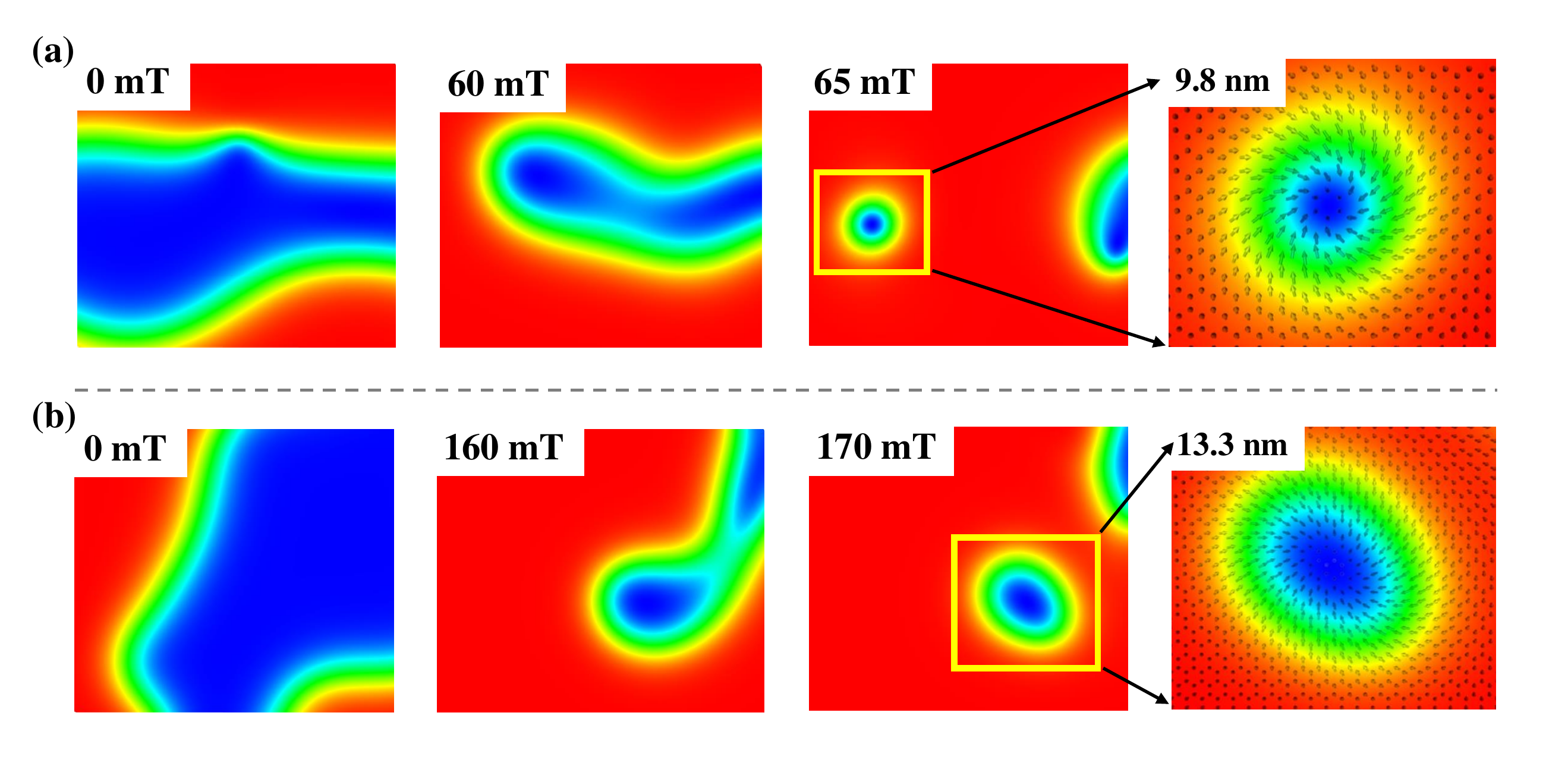}
		\caption{\label{fig5}
			Spin textures of Cr$_2$I$_3$Cl$_3$ at strains of (a) $0\%$ and (b) $2\%$ under different applied magnetic fields.
		}
\end{figure*}

With the magnetic parameters obtained from a first-principles approach, we perform micromagnetic simulations with the Landau--Lifshitz--Gilbert (LLG) equation~\cite{RN1984,RN1985} in the framework of \textit{Spirit}~\cite{RN1055}.
As shown in Fig.~\ref{fig5}(a), when no magnetic field is applied, we obtain wide domains separated by N\'eel-type domain walls. Then, with a small magnetic field of 60~mT, the spin-down domain shrinks to a wormlike one. When the magnetic field is increased further  to 65~mT, an isolated skyrmion is induced, with a small diameter of 9.8~nm. Figure~\ref{fig5}(b) shows the evolution of spin textures in Cr$_2$I$_3$Cl$_3$ under different applied magnetic fields at a strain of $2\%$. Similar to the case at a strain of $0\%$, a small-sized skyrmion with a diameter of 13.3~nm can be induced by a magnetic field of 170~mT. Such a small skyrmion ($\sim$10~nm) is technologically desirable, since it can significantly enhance the storage density of skyrmion-based next-generation information memory devices~\cite{RN2003,RN3705}.

\section{Conclusions}
\label{IV}

To sum up, by first-principles calculations and micromagnetic simulations, we have investigated in detail the magnetic parameters and spin textures of Janus Cr$_{2}$X$_{3}$Y$_{3}$ monolayers under biaxial strain. We have found that the DMI and MAE can be significantly enhanced by tensile strain. With a compressive strain of $-2\%$, a chirality reversal of DMI and a switch of MAE from off-plane to in-plane appear in Cr$_2$Cl$_3$Br$_3$ and Cr$_2$I$_3$Cl$_3$, respectively. We have also explored the microscopic physical mechanisms of DMI and MAE in  view of the strong SOC induced by the heavy nonmagnetic halogen atoms. In particular, we have explained the mechanisms of the peculiar magnetic transition in the Cr$_2$I$_3$Cl$_3$ monolayer in terms of  direct exchange and superexchange interactions. Moreover,  in the unstrained Cr$_2$I$_3$Cl$_3$ monolayer, an isolated skyrmion with sub-10~nm diameter has been found which is desirable for spintronic applications. Our work has enlarged the family of 2D Janus materials, as well as providing guidance for further research on the DMI and chiral spin textures.

\begin{acknowledgments}
This work was supported by the National Natural Science Foundation of China (No.~11804301), the Natural Science Foundation of Zhejiang Province (No.~LY21A040008), and the Fundamental Research Funds of Zhejiang Sci-Tech University (No.~2021Q043-Y).
\end{acknowledgments}

\bibliography{ref}
\end{document}